\documentclass[amsmath,amssymb,floatfix]{nature}
\usepackage{hyperref}
\usepackage{times}
\usepackage{graphicx}
\usepackage{dcolumn} %
\usepackage{bm}      %
\usepackage{subfigure}
\usepackage[usenames]{color}
\usepackage{rotating}
\usepackage{fontenc}
\usepackage{amsthm}
\usepackage{hyperref}
\usepackage{cancel}
\usepackage{ulem}
\usepackage{array}
\usepackage{amssymb}
\usepackage{amsfonts}
\usepackage{amsmath}
\usepackage{mathrsfs}

\definecolor{darkred}{RGB}{139,0,0}
\definecolor{chartreuse}{RGB}{127,255,0}
\definecolor{goldenrod}{RGB}{218,165,32}
\definecolor{gray}{RGB}{127,127,127}

\definecolor{Magenta}{RGB}{255, 0,255}
\definecolor{Orange}{RGB}{255,165, 0}
\definecolor{Gray}{RGB}{127,127,127}

\newcommand{\be}{\begin{equation}}
\newcommand{\ee}{\end{equation}}
\newcommand{\bea}{\begin{eqnarray}}
\newcommand{\eea}{\end{eqnarray}}
\newcommand{\bw}{\begin{widetext}}
\newcommand{\ew}{\end{widetext}}
\newcommand{\mm}{\mathrm}

\newcommand{\bi}{\begin{itemize}}
\newcommand{\ei}{\end{itemize}}

\title{Prospect Theory for Online Financial Trading}
\author{Yang-Yu Liu$^{1,\ast}$, Jose C. Nacher$^{2,\ast}$, Tomoshiro
  Ochiai$^{3,\ast}$, Mauro Martino$^{4}$, Yaniv Altshuler$^{5}$}

\begin{document} 
\maketitle \\ \date{\hspace{0.1in}\scriptsize{\today}\\ $\ast$  
These authors contributed equally to this work.}

\begin{affiliations}
\item{Channing Division of Network Medicine, Brigham and Women's
Hospital, Harvard Medical School, Boston, Massachusetts 02115, USA}
\item{Department of Information Science, Faculty of Science, Toho University, Miyama 2-2-1, Funabashi, Chiba 274-8510, Japan.}
\item{Department of Social Information Studies, Otsuma Women's University, 2-7-1 Karakida,
Tama-shi, Tokyo 206-8540, Japan.}
\item{Center for Innovation in Visual Analytics, Watson Research Center, IBM, Cambridge, Massachusetts 02142, USA.}
\item{MIT Media Lab, Cambridge, Massachusetts 02139, USA.}
\end{affiliations}

\begin{abstract}
Prospect theory is widely
viewed as the best available descriptive model of how people evaluate
risk in experimental
settings\cite{Kahneman-79,Tversky1981,Tversky-92,Wakker93,Schmidt08,Barberis13}. %
According to prospect theory, people
are typically risk-averse with respect to gains and risk-seeking with respect to losses, known as the ``reflection
effect''. 
People are usually much more sensitive to losses than to
gains of the same magnitude, a phenomenon called ``loss
aversion''\cite{%
Schmidt02c,Schmidt02d,Schmidt05,Brooks05,Abdellaoui-07,Schmidt08b}.
Despite of the fact that 
prospect theory has been well developed in behavioral economics at the
theoretical level, there exist very few large-scale empirical studies and most of
them have been undertaken with micro-panel
data\cite{Camerer-98,Abdellaoui00,Wu05,Abdellaoui-07,Zhang-09,Booij10,Abdellaoui2011}. Here
we analyze 
over 28.5 million trades made by 81.3 thousand traders of an online financial
trading community over 28 months, aiming to explore the large-scale
empirical aspect of prospect theory. By analyzing and comparing the
behavior of winning and losing trades and traders,   
we find clear evidence of the reflection effect and the loss aversion
phenomenon, which are essential in prospect theory. This work hence demonstrates an unprecedented
large-scale empirical evidence of prospect theory, which has immediate
implication in financial trading, e.g., developing new trading
strategies by minimizing the impact of the reflection effect and the
loss aversion phenomenon. 
Moreover, we introduce three novel behavioral metrics %
to differentiate winning and losing traders %
based on their historical trading behavior. This offers us potential
opportunities to augment online social trading where traders are allowed
to watch and follow the trading activities of others\cite{Pan2013}, by predicting
potential winners %
based on their historical trading behavior rather
than their trading performance at any given point in time.  
\end{abstract}

We live life in the ``big data'' era. Many of our daily activities
such as checking emails, making mobile phone calls, posting blogs on
social media, shopping with credit cards and making financial trading
online, will leave behind our digital traces of various kinds that can
be compiled into comprehensive pictures of our behavior.  
The sudden influx of data is transforming social sciences at an
unprecedented pace~\cite{Lazer-Science-09,Battiston12,Preis13}. Indeed, researchers are 
moving in a few years from dealing with interviews of a few dozens of
people by crafting survey questionnaire to experiments
involving millions of subjects using social media~\cite{Bond-Nature-12}.

The availability of huge amounts of digital data 
also prompts us to rethink some fundamental perspectives of complex human
behavior.  
In this work we focus on economic decision under
risk, a key subject of behavior economics~\cite{Diamond-07}. Successful behavior economic
theories acknowledge the complexity of human economic behavior and  
introduce models that are well grounded in
psychological research. 
For example, prospect theory is viewed as the best available descriptive
model of how people evaluate
risk~\cite{Kahneman-79,Tversky1981,Tversky-92}. Prospect theory states that people make 
decisions based on the potential value of losses and gains rather than
the final outcome, and that people evaluate these losses
and gains using certain heuristics. 
Despite the fact that prospect theory offers many remarkable
insights and has been studied for more than three
decades, there exist very few large-scale empirical
research and relatively few well-known and broadly accepted
applications of prospect theory in economics and
finance\cite{Barberis13}. 
The emergence of online social trading platforms and the
availability of burgeoning volume of financial transaction data of
individuals help us explore the empirical aspect of prospect theory to
an unprecedented large-scale. 
Moreover, analyzing
  the trading behavior at the individual level offers an excellent opportunity to
  develop pragmatic financial applications of prospect theory.

By harnessing the wisdom of the crowds to our benefit and
gain, social trading has been a revolutionary way to approach
financial market investment. 
Thanks to various Web 2.0 applications, nowadays online traders can
rely on trader generated financial content as the major information
source for making financial trading decisions. 
This ``data deluge'' raises some new questions, answers to
which could further deepen our understanding of the complexity of
human economic behavior and improve our social trading experience.  
For example, many social trading platforms allow us to follow top
traders, known as gurus or trade leaders, and directly invest our money like they do. The
question is then how to identify those top traders. Analyzing
their historical trading behavior would be a natural starting 
point. 
The financial transaction data used in this work comes from 
an online social trading platform for foreign
exchanges and commodities trading. 
This trading platform allows traders to take both long and short
positions, with a minimal bid of a few dollars as well
as leverage up to 400 times. 
The most important feature of social trading platform is that each trader
automatically has all trades uploaded to the platform where trades can
be displayed in a number of statistical ways, such as by the amount of profit
made. 
Traders can then set their accounts to copy one or more
trades made by any other traders, in which case the social trading
platform will %
automatically execute the trade(s).
Accordingly, there are three types of trades%
: (1) Single (or
non-social) trade: Trader A places a normal trade by himself or herself; (2) Copy trade: Trader A places
exactly the same trade as trader B's one single trade; (3) Mirror trade: Trader A
automatically executes trader B's every single trade, i.e., trader A
follows exactly trader B's trading activities. Both (2) and (3) are
hereafter referred to as social trading.

There are about 3 million registered %
accounts in this online social trading platform. 
Some of them are
practice accounts, i.e., trading with virtual money. Our data are
composed of over 28.5 million trades made by 81.3 
thousand traders %
trading with real money from June 2010 to October 2012.
There are 31.8\% single trades, 0.6\% copy trades
and 67.6\% mirror trades. Apparently, social trading dominates over 
this trading platform during the time window of our data. 
It will be desirable to learn how to select the best traders to
follow so that we can further improve our social trading experience
--- a pragmatic motivation of our current work. Quantitatively
analyzing trading activities of traders within the framework of
behavior economics naturally fits the goal. Ultimately, we would like
to be able to predict potential winners based on their historical trading
behavior so that we can take full advantage of the social trading
paradigm. %

We first need to demonstrate if social trades really help. 
In Fig.~1 we compare the fraction of winning
trades ($N_+/N$) and return on interest (ROI (\%) $:= \mm{net
profit}/\mm{investment}  \times 100$) of the three different trade
types. 
We find that all three trade types have more than 50\% chance
to generate positive net profit (see
Fig.~1a). Among them, mirror trade has the 
highest chance ($\approx 83\%$), much higher than that of single or
copy trade. This indicates that in average social trades (especially
mirror trades) indeed help traders win more frequently than non-social 
trades.  
Interestingly, not all the trade types have positive
average ROI (see Fig.~1b).
In fact, only mirror trade has positive average ROI ($\approx 0.03\%$), i.e., it
generates profit, consistent with previous
results~\cite{Pan2013}. In terms of ROI, social trades do not necessarily perform better
than non-social trade. We notice that copy trade even has higher negative ROI
than non-social trade, which simply implies that copying someone based
on past performance can be dangerous.

Overall, mirror trade outperforms both single and copy
trades. Yet, the better performance of mirror
trade comes at the price that its winning trades have much lower ROI
($\approx 0.177\%$) than that of copy and single trades (see
Fig.~1c); while its losing trades have
significantly higher negative ROI ($\approx -0.9\%$) than that of 
copy and single trades (see Fig.~1d).  
In other words, mirror trade typically does not generate high profit
for winning positions but generate high loss for losing
positions. Since mirror trade has very high chance of winning, %
the average ROI of mirror trade turns out to be positive. 
This implies that there is still much room to improve our social
trading experience. 

To further understand the difference between social and non-social
trade types, we calculate their duration distributions $P(\tau)$ (see Fig.~2). Here the
duration $\tau$ of a trade or position is defined to be its holding time (in unit
of millisecond),
i.e., $\tau:= t_\mm{closed}-t_\mm{opened}$ where $t_\mm{opened}$ and
  $t_\mm{closed}$ are the position opened and closed time,
  respectively.  
Interestingly, $P(\tau)$ displays similar fat-tailed distribution for all
different trade types. There are very few positions that were held for
very long time (more than one month). Most of the positions were held
for less than half an hour. We also notice that for losing positions,
many of them are held for less than one second, while for winning
positions they are most likely held for longer than one second. 
This might be due to the so-called bid-ask spread. The price we can sell
(bid) and the price we can buy (ask) is different at each time
point. It is almost impossible for traders to overcome the spread within very short
holding time interval (e.g., one second) by using online financial
trading platforms. %
For $\tau \in [10^3, 10^5]$, we find that for all different trade types $P(\tau)$ of
positive trades is much lower than that of negative trades, i.e.,
winning probability is much less than 50\% in this particular holding time window.
Although the duration distributions of different trade types share
many similar features, we do observe a noticeable difference in the regime of $\tau >
10^5$, i.e., longer than one minute. 
We find that for non-social trades with $\tau > 10^{5}$, $P(\tau)$ of
positive trades is roughly the same as that of negative trades. In
other words, if the holding time of a non-social trade is longer than
one minute, the winning chance is about 50\%.  
For copy trades with $\tau \in (10^{5}, 10^{8})$, $P(\tau)$ of
positive trades is slightly higher than that of negative trades. 
For mirror trades with $\tau \in [10^{5}, 10^9]$, $P(\tau)$ of
positive trades is significantly higher than that of negative trades.  
In other words, if the holding time of a mirror trade is longer than
one minute and less than one week, then its winning chance is much
higher than 50\%. 

We also calculate the trade duration as a function of the
net profit for different trade types (see Fig.~3). 
We draw the box-and-whisker plot of duration ($\tau$)
for trades with net profit ($p$) binned logarithmically. (For negative
trades $p<0$, we bin them using $|p|$.) 
We denote the median value of durations as $\tau_\mm{m}$.  
We find that for all trade types $\tau_\mm{m}$ shows asymmetric
behavior: $\tau_\mm{m}$ of losing positions with loss $-p$ is
generally higher than that of winning positions with profit $p$.
This is a reflection of the so-called ``disposition effect'':
investors tend to sell financial assets whose price has 
increased while keeping asserts that have dropped in
value~\cite{Shefrin1985,Weber1998,Barberis2009}. 
In other words, investors are less willing to recognize losses, but are
more willing to recognize gains. 
This is a typical irrational behavior that can be partially explained by the  ``loss aversion''
phenomenon and the ``reflection effect'' in prospect theory. %
We also notice clear differences between mirror trade and the other
two trade types. 
For both non-social and copy trades $\tau_\mm{m}$ generally increases
as $|p|$ increases in either positive or negative direction, and $\tau_\mm{m}$
of losing trades increases much faster as $|p|$ increase than
$\tau_\mm{m}$ of winning trades increases as $p$ increase.
While for mirror trade, $\tau_\mm{m}$ increases very slowly as $p$
increases for positive positions. For mirror trades with negative $p$,
$\tau_\mm{m}$ increases initially as $|p|$ increases, but quickly
reaches a plateau.
In other words, the disposition effect is lessened in mirror
trade. 
It has been shown that %
more experienced investors are less affected by
the disposition effect~\cite{Costa2013}. This might explain the good performance of
mirror trade.  
To characterize the trading behavior of traders and identify potential
trade leaders, we introduce four %
  behavioral metrics: 
(1) Risk-reward ratio $r:= \frac{\langle p_\mm{+} \rangle}{\langle
  |p_\mm{-}| \rangle}$, where  $\langle p_\pm \rangle$ represents the
average profit of positive/negative trades made by each trader. $r>1$
means that traders in average gain more in positive trades than the loss
in negative trades. 
(2) Win-loss holding time ratio $s := \frac{\langle \tau_+
  \rangle}{\langle \tau_-\rangle}$, where $\langle \tau_\pm \rangle$
represents the average holding time of positive/negative trades made
by each trader. $s>1$ means that traders in average hold positive position
longer than negative position. 
(3) Win-loss ROI ratio $u:= \frac{\langle \mm{ROI}_+
      \rangle}{\langle |\mm{ROI}_-| \rangle}$, where 
$\langle \mm{ROI}_\pm \rangle$ represents the
average ROI of positive/negative trades made by each trader. $u>1$
means that traders in average have larger absolute ROI in positive
trades than in negative trades. 
(4) Winning percentage $w := \frac{N_+}{N_++N_-}$,
where $N_\pm$ represents the number of positive/negative trades made
by each trader. %
$w>1/2$ simply means that traders in
  average have larger chance of winning than losing a trade.%

Note that if all traders trade pure randomly without any human emotions we would expect that the
distributions of all the four metrics show symmetric behavior around
$r=1$, $s=1$, $u=1$ and $w=1/2$, respectively. 
Yet, in reality traders behave quite differently from random (see
Fig.~4). 
We find the risk-reward distribution $P(r)$ displays strongly asymmetric behavior around $r=1$
(black dotted line in Fig.~4a). For $r>1$, $P(r)$ follows a power
law over almost 3 decades, which means it is extremely difficult to find traders with very large $r$; while
for $r\le 1$, $P(r)$ is almost a constant, which means traders with
$r\le 1$
are almost uniformly distributed. 
By splitting the traders into two groups: winning and losing traders
(i.e., traders with final net profit or net loss) and
calculating their $P(r)$ with appropriate normalization based on the
fractions of winning and losing traders ($14.7\%$ and $85.2\%$,
respectively), we find that the two groups behave in drastically
different ways  (see Fig.~4b). Winning traders' $r$-range
spans over $[10^{-2}, 10^3]$; while losing traders' $r$-range is given by
$[10^{-4}, 10^1]$. The uniform $P(r)$ for $r<1$ is largely due to
losers; while the power-law of $P(r)$ for $r>10$ is purely due to
winning traders. 
We also notice that
for $r<r^*=4$ (pink line), $P(r)$ of losing traders is significantly higher than that of
winning traders; while for $r>r^*$, it is in the opposite case.

We find the win-loss duration ratio distribution $P(s)$ also displays
strongly asymmetric behavior around $s=1$ (black dotted line in
Fig.~4c). For $s>1$, $P(s)$ almost follows a power
law over 5 decades, which means it is extremely difficult to find
traders with very large $s$; while for $s\le 1$, $P(s)$ decays very
slowly as $s$ decreases, which means traders with $s\le 1$ are almost
uniformly distributed. This indicates that most traders hold losing
positions for a longer time than winning position, a typical
disposition effect.  
Comparing winning and losing traders' $P(s)$ is also interesting (see
Fig.~4d). Though their $s$-ranges are almost the
same, we notice that for $s<s^*=100$ (pink line), $P(s)$ of losers is
significantly higher than that of winners; while for $s>s^*$, it is in
the opposite case.

The win-loss ROI ratio distribution $P(u)$ shows a strong peak around
$u=0.2$ and a strongly asymmetric behavior around $u=1$ (black dotted
line in Fig.~4e). For $u>0.2$, $P(u)$ follows a power
law over 3 decades, which means it is extremely difficult to find
traders with very large $u$. Interestingly, for $u\le 0.2$, $P(u)$
also follows a power-law over almost 2 decades.
We find $P(u)$'s of winning and losing traders are also very different (see
Fig.~4f). Winning traders' $u$-range
spans over $[10^{-1}, 300]$; while losing traders' $u$-range is given by
$[5\times 10^{-4}, 300]$. For $u<u^*=2$ (pink line), $P(u)$ of losing
traders is significantly higher than that of winning traders; while
for $u>u^*$, it is in the opposite case. For $u<0.06$, almost all
traders are losing traders.

Note that a large portion of traders ($85.2\%$) are
losing traders with final net loss. The fact that those
losing traders typically have $r<1, s<1$ and $u<1$ could be
explained by the  ``loss aversion'' phenomenon and the ``reflection
effect'' in prospect theory as
follows. For positive positions,  traders tend to be
risk-averse and will close the position quickly to take a small profit
or a small ROI. 
While for losing positions, traders tend to be risk-seeking and
reluctant to close the positions as quick as they should. Instead they
keep waiting and hoping to recover the
loss.  If indeed this happens, then they become risk-averse and tend
to close the position quickly to take a small profit and result in a
small ROI. If unfortunately this does not happen, they waste not only
valuable time but also a lot of money, rendering large negative ROI. 
Thus, the losing traders have to suffer from their irrational trading
behavior %
that can be described by prospect theory.

One may naively consider the winning percentage type of behavioral metrics will
help us identify gurus. Here we show it is not the
case. Fig.~4g displays the winning percentage
distribution $P(w)$, which is asymmetric around $w=1/2$. (Note that
$P(w)$ has significant peaks around some \sout{simple} rational numbers $w=0,
1/2, 1/3, 2/3, ...,1$, which are due to %
traders who made very few transactions.) %
Again, we find winning and losing traders' $P(w)$ show dramatically
different behavior (see Fig.~4h). For $w<w^*=0.95$,
$P(w)$ of losing traders is significantly higher than that of 
winning traders; while for $w>w^*$ winning traders dominate.
The value of $w^*$ is so high that using it to select trade leaders is
almost infeasible. 
We also notice that for $w<w^*$, losing traders actually dominate for
a wide range of $w$. Yet, they are still losing money eventually due
to their very low risk-reward ratio $r$. In other words, they win many
times with small positive profit, but once they lose they lose a lot.  

In principle large values of those metrics do not imply net profit at
all. For example, traders  with $r \gg 1$ could be simply 
due to a few trades with very large profit, but many trades
with very small loss. 
Yet, the above analysis yields three characteristic values ($r^*=4,
s^*=100, u^*=2$) that can be used to statistically predict potential
winning traders with high probability. 
We admit that those characteristic values may slightly depend on
the particular dataset or trading platform. We emphasize that the
strategy of using characteristic values of novel behavioral metrics %
to identify potential winning traders should be
applicable to general social trading platforms.
Furthermore, the existence of characteristic values ($r^*$, $s^*$, $u^*$) of these behavioral
metrics indicates the importance of controlling human emotion to
minimize the impact of the reflection effect and the loss aversion
phenomenon for better trading performance.    
\section*{Discussion}

The dynamics of financial trading is governed by individual human
decisions, which implies that the trading performance could be
significantly improved by understanding better the underlying human
behavior. In this work we systematically analyze over 28.5 million
trades made by 81.3 thousand traders of an online financial trading
platform. By analyzing and comparing the performance of social and
non-social trades, winning and losing traders, we find clear
evidences of the reflection effect and the loss aversion phenomenon, which are essential in
prospect theory of behavior economics. 
Many losing traders have very small risk-reward ratio ($r$), win-loss
duration ratio ($s$), and win-loss ROI ratio ($u$), suggest that we should develop
new trading strategies by systematically minimizing the impact of the
reflection effect and the loss
aversion phenomenon, e.g., through intentionally controlling $s, u$ and/or $r$ to
fight over our human nature and rationalize our trading behavior.   
To provide traders many preferences in discovering gurus or trade leaders, social
trading platforms rank traders on many different metrics,
e.g., popularity (number of followers), profitable weeks, and the
personal return of rate calculated from the modified Dietz 
formula~\cite{Dietz2004}, etc. Those different metrics typically yield
different ranking lists, which effectively renders choosing the gurus 
something of an inspirational affair or a delicate trick.   
Moreover, traders should take into account the risks taken by these
gurus in order to obtain the returns that they make. Unfortunately,
not all the metrics rank the performance of traders on a risk-adjusted
basis. 
Here we propose three novel behavioral metrics (risk-reward ratio $r$, win-loss holding
time ratio $s$, and win-loss ROI ratio $u$), which reveal the essence
of 
prospect theory, the best available descriptive model of how people
evaluate risk in behavioral economics. 
These metrics are defined for each trader by comparing his/her typical
behavior in winning and losing trades, and hence are naturally
risk-adjusted. 
Our analysis suggests that these metrics can be used to statistically
predict potential winning traders, offering pragmatic opportunities to
further improve our social trading experience. 

\begin{addendum}

 \item[Author Contributions] 
   Y.-Y.L, J.N. and T.O. conceived and executed the research, and
   contributed equally to this project. M.M. and Y.A. provided the raw
   data and involved in valuable discussions. Y-Y.L wrote the manuscript
   with the assistance of J.N.,T.O., M.M. and Y.A.
 
 \item[Competing Interests] The authors declare that they have no
   competing financial interests.
   
  \item[Correspondence] Correspondence and requests for materials should be
    addressed to Yang-Yu Liu ~(email: yyl@channing.harvard.edu). 
  \end{addendum}

\newpage
\begin{figure}
\begin{center}
\includegraphics[width=0.48\textwidth]{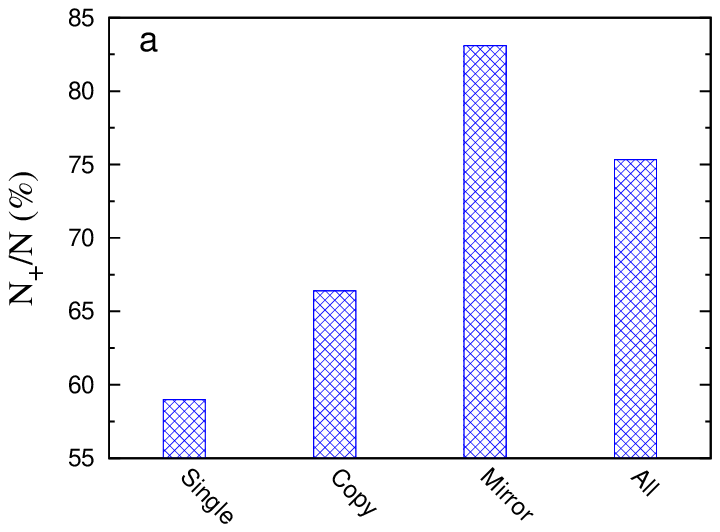}
\includegraphics[width=0.48\textwidth]{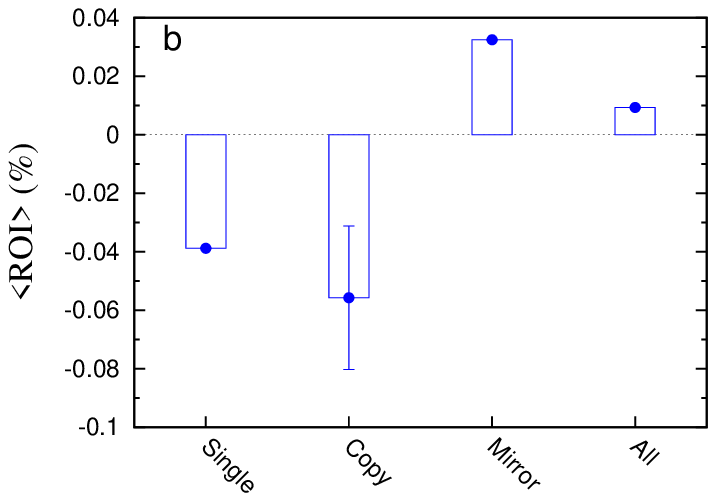}
\includegraphics[width=0.48\textwidth]{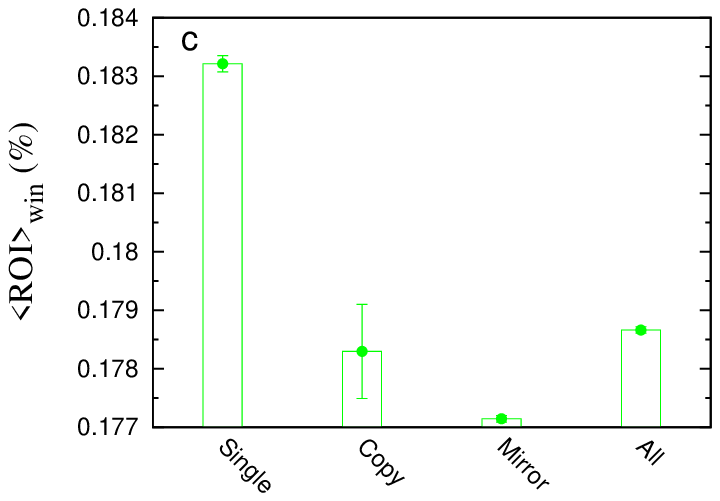} 
\includegraphics[width=0.48\textwidth]{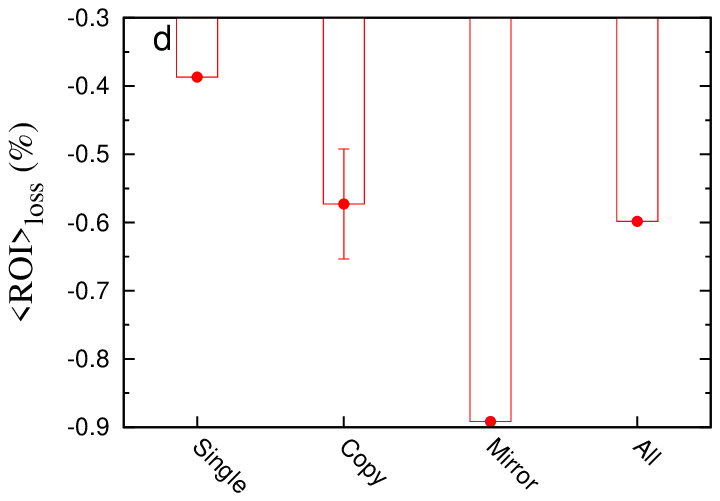} 
\end{center}
\caption{{\bf Performance comparison of different types of trades.}
{\bf (a)} Fraction of positive trades. Mirror trade has the highest fraction
of positive trades. 
{\bf (b)} Mean ROI. Mirror trade is the only trade type that has the positive
$\langle \mm{ROI} \rangle$. Here error bars mean the standard error of
the mean (SEM).
{\bf (c)} Mean ROI of positive trades. Mirror trade has the lowest $\langle \mm{ROI} \rangle$ for positive trades. 
{\bf (d)} Mean ROI of negative trades. Mirror trade has the highest
negative $\langle \mm{ROI} \rangle$ for negative trades. }\label{fig:ROI_tradetype}
\end{figure}Fig.1

\newpage
\begin{figure}
\begin{center}
\includegraphics[width=0.48\textwidth]{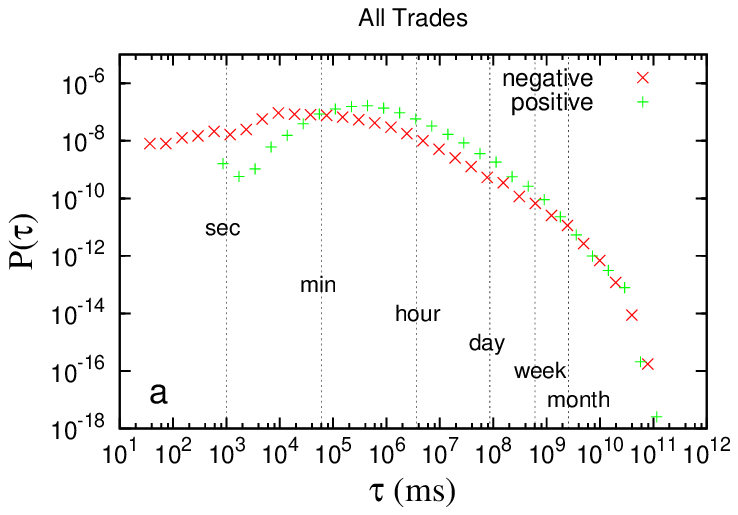} %
\includegraphics[width=0.48\textwidth]{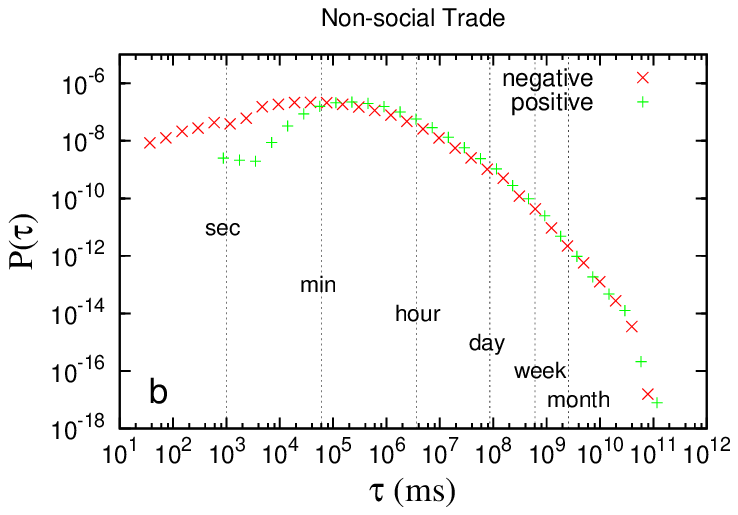} %
\includegraphics[width=0.48\textwidth]{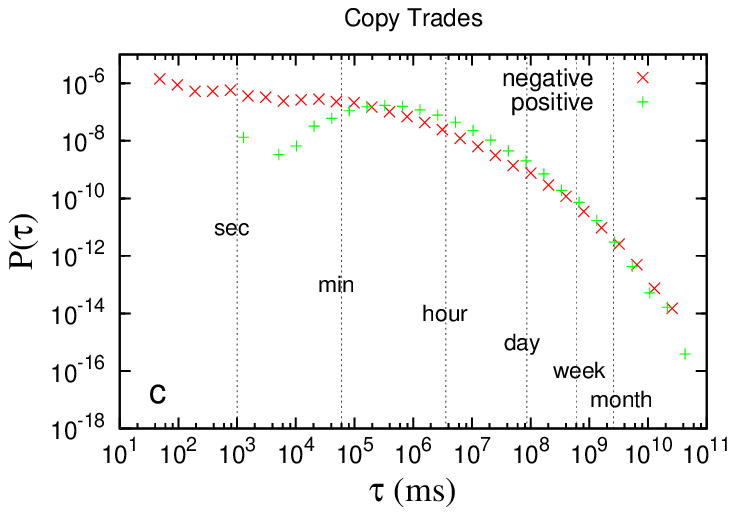} %
\includegraphics[width=0.48\textwidth]{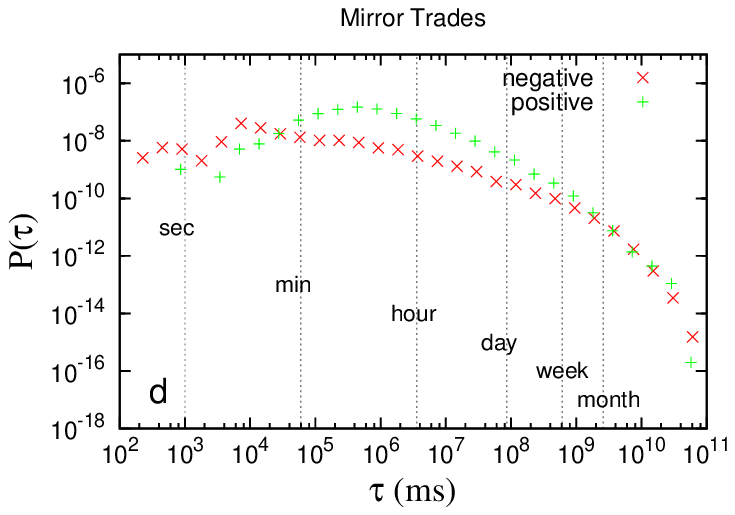} %
\end{center}
\caption{{\bf Duration distribution of different trade types.}
 For each trade type, we further distinguish negative and positive trades
  based on their net profit. The trades with zero net profit are
  negligible. The duration distributions of negative and positive
  trades are normalized according to their corresponding occurrence.     
{\bf (a)} All trades.
{\bf (b)} Non-social trades.
{\bf (c)} Copy trades.
{\bf (d)} Mirror trades.}
\label{fig:P_tau}
\end{figure}
Fig.2

\newpage
\begin{figure}
\begin{center}
\includegraphics[width=0.48\textwidth]{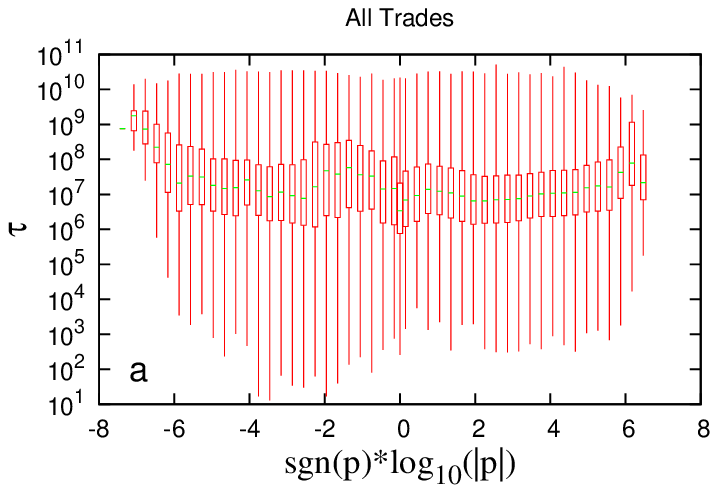} 
\includegraphics[width=0.48\textwidth]{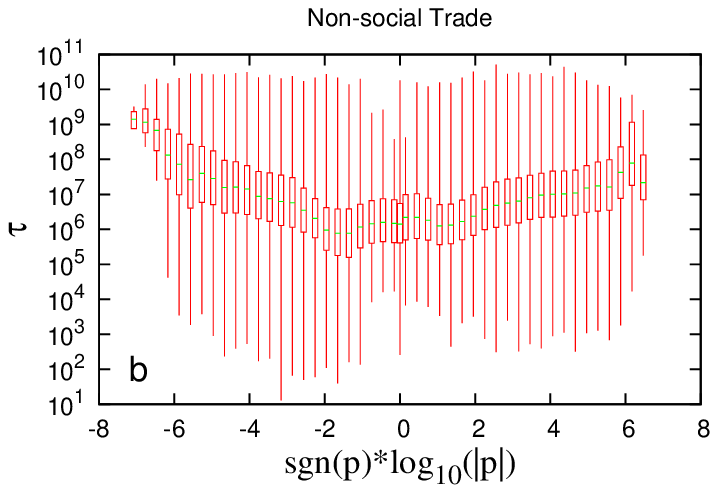} 
\includegraphics[width=0.48\textwidth]{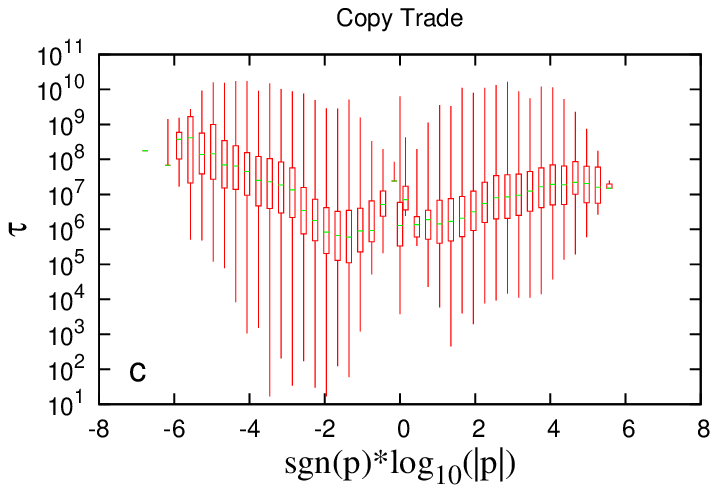} 
\includegraphics[width=0.48\textwidth]{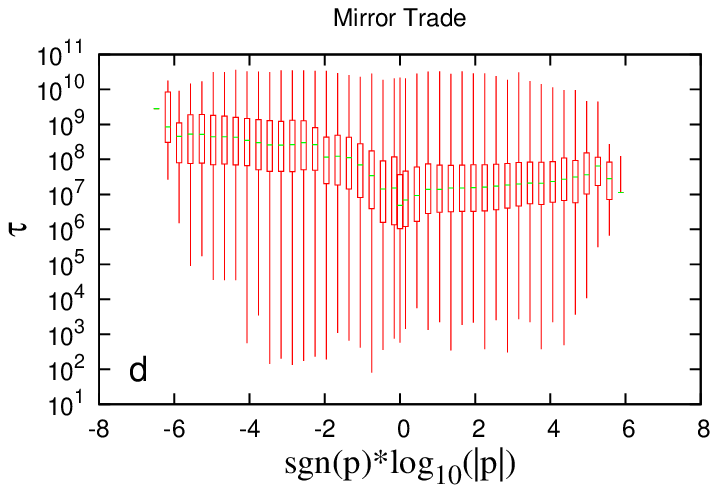} 
\end{center}
\caption{%
{\bf Disposition effect in different trade types.}
Here, we bin the net profit $p$ of different trade types in logarithmic bins. (If $p<0$, we bin it
using $|p|$.) For the trades contained in each bin, we draw the box-and-whisker plot for
their duration ($\tau$), representing the minimum, first
quartile, median, third quartiles, and maximum of the data in the bin.  
{\bf (a)} All trades.
{\bf (b)} Non-social trades.  
{\bf (c)} Copy trades.
{\bf (d)} Mirror trades. 
}\label{fig:d-p}
\end{figure}Fig.3

\newpage
\begin{figure}
\begin{center}
\includegraphics[width=0.43\textwidth]{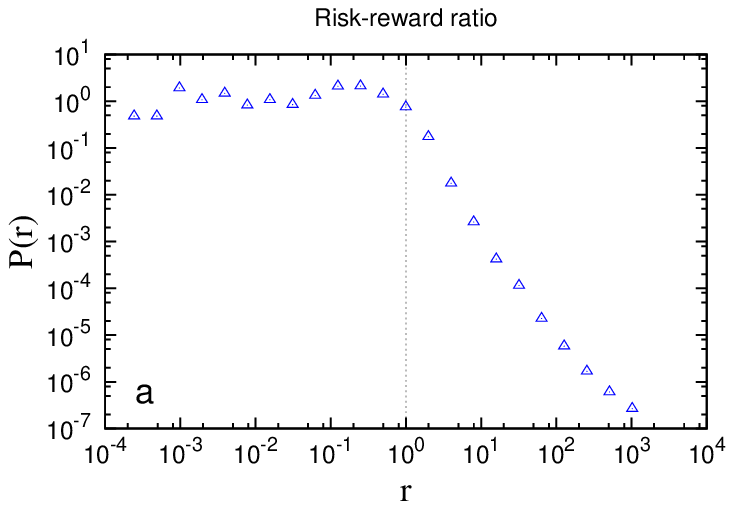}
\includegraphics[width=0.43\textwidth]{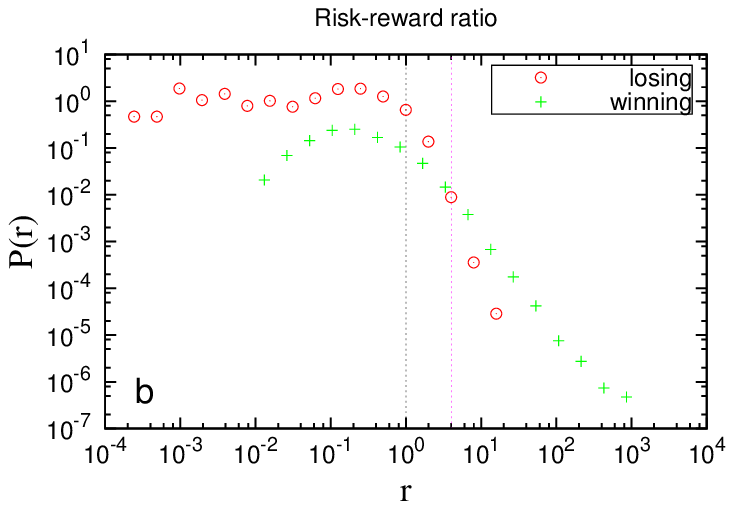}
\includegraphics[width=0.43\textwidth]{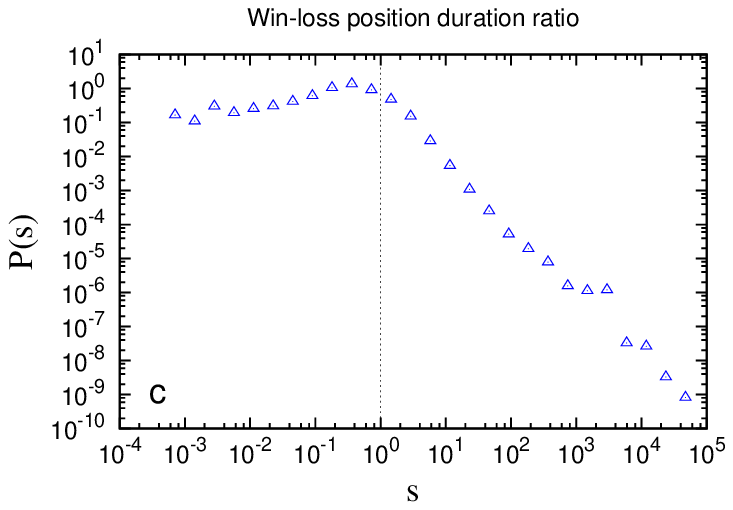}
\includegraphics[width=0.43\textwidth]{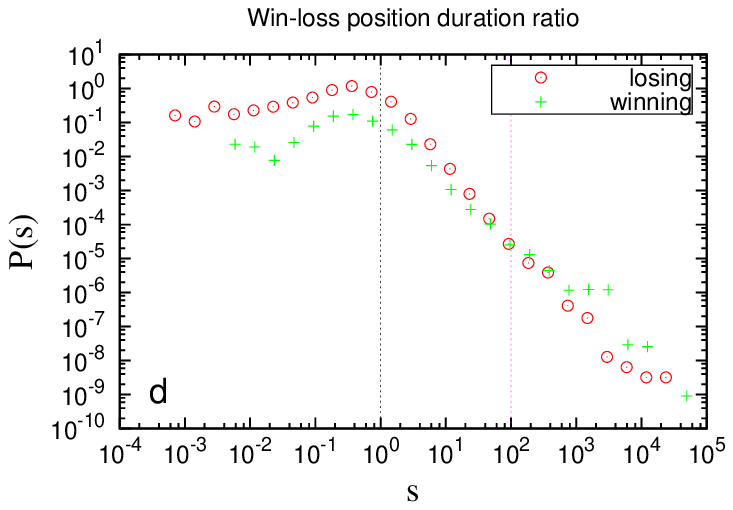}
\includegraphics[width=0.43\textwidth]{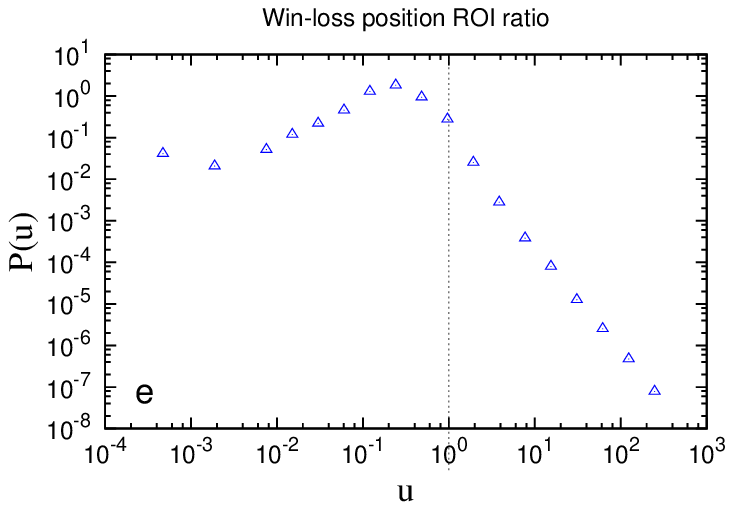}
\includegraphics[width=0.43\textwidth]{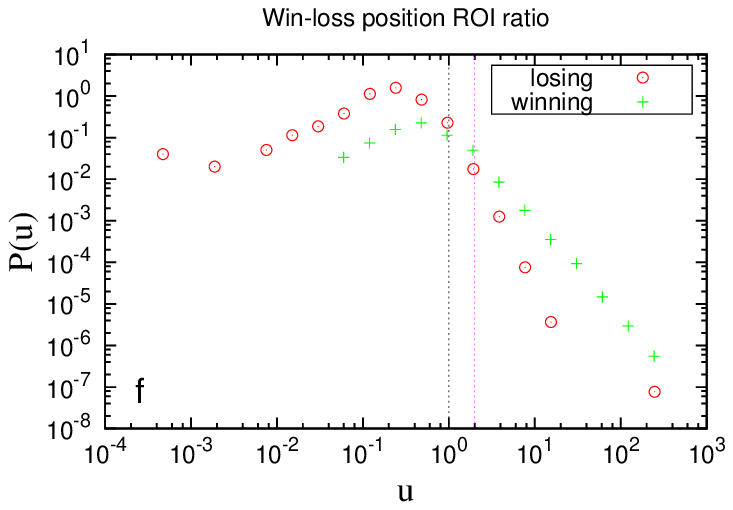}
\includegraphics[width=0.43\textwidth]{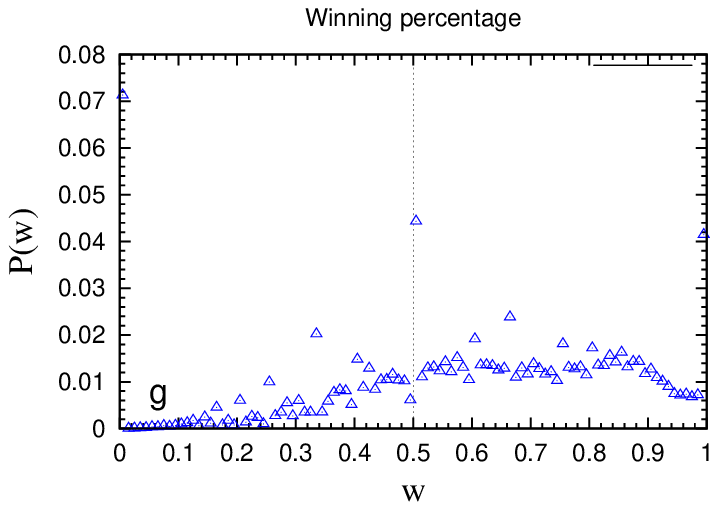}
\includegraphics[width=0.43\textwidth]{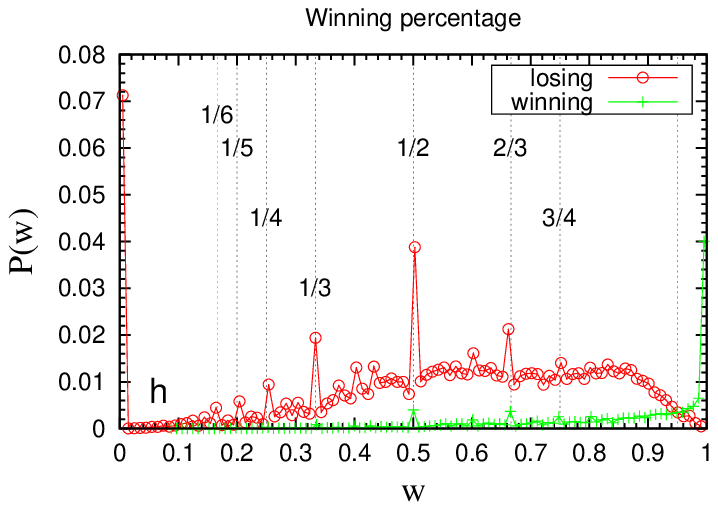}
\end{center}
\caption{{\bf Characterizing winning and losing traders based on
    historical trading behavior.} 
{\bf (a, b)} Distribution of risk-reward ratio ($r:= \frac{\langle p_\mm{+}
      \rangle}{\langle |p_\mm{-}| \rangle}$).
$p_\mm{+}$ and $p_\mm{-}$ are the profit of winning positions and the loss
of losing positions, respectively, of traders.
{\bf (c, d)} Distribution of win-loss waiting time ratio
    ($s := \frac{\langle \tau_+ \rangle}{\langle
      \tau_-\rangle}$). Here $\langle \tau_+ \rangle$ and $\langle
    \tau_- \rangle$ are the average duration time of winning and
    losing positions, respectively, of traders.  
{\bf (e, f)} Distribution of win-loss ROI ratio ($u:= \frac{\langle \mm{ROI}_+
      \rangle}{\langle |\mm{ROI}_-| \rangle}$).
$\mm{ROI}_\mm{+}$ and $\mm{ROI}_\mm{-}$ are the ROI of winning and
losing positions, respectively, of traders.
{\bf (g, h)} Distribution of winning percentage ($w :=
\frac{N_+}{N_++N_-}$) of traders. 
$N_+$ and $N_-$ are the number of winning and
losing positions, respectively, of traders.
}\label{fig:P_r_s_u_w}
\end{figure}Fig.4
\newpage

\end{document}